\documentstyle[aps,preprint]{revtex}

\begin{document}
\author{Remo Garattini\thanks{%
Contributed paper to the IX Marcel Grossmann Meeting (Rome, July 2000)%
\newline
To appear in the Proceedings (World Scientific, Singapore) }}
\title{Wormholes and Spacetime Foam:\\
an approach to the cosmological constant and entropy}
\address{Facolt\`a di Ingegneria, Universit\`a di Bergamo,\\
Viale Marconi 5, 24044 Dalmine (Bergamo), Italy\\
E-mail:\\
Garattini@mi.infn.it}
\maketitle

\begin{abstract}
This paper summarizes the contribution presented at the IX Marcel Grossmann
Meeting (Rome, July 2000). A simple model of spacetime foam, made by $N$
Schwarzschild wormholes in a semiclassical approximation, is here proposed.
The Casimir-like energy of the quantum fluctuation of such a model and its
probability of being realized are computed. Implications on the
Bekenstein-Hawking entropy and the cosmological constant are considered. A
proposal for an alternative foamy model formed by $N$ Schwarzschild-Anti-de
Sitter wormholes is here considered.
\end{abstract}

\newpage

The term Spacetime Foam was used for the first time by J.A. Wheeler to
indicate that spacetime may be subjected to quantum fluctuations in topology
and metric at the Planck scale \cite{1}. Since then a lot of work has been
done, especially in the direction of string theory and of Euclidean Quantum
Gravity, to see if a nontrivial vacuum can be considered for a quantum
theory of gravitation. In this paper we will consider a different approach
based on a proposed model of spacetime foam, made by $N$ coherent
Schwarzschild wormholes \cite{2}. The main reason which has led to consider
such a model is based on a Casimir energy computation described by the
following Hamiltonian
\begin{equation}
\Delta E\left( M\right) =E\left( M\right) -E\left( 0\right) =\frac{%
\left\langle \Psi \left| H_{\Sigma }^{Schw.}-H_{\Sigma }^{Flat}\right| \Psi
\right\rangle }{\left\langle \Psi |\Psi \right\rangle }+\frac{\left\langle
\Psi \left| H_{ql}\right| \Psi \right\rangle }{\left\langle \Psi |\Psi
\right\rangle }.
\end{equation}
The first term represents the difference of the Hamiltonians evaluated on
the Schwarzschild and flat spacetime respectively and the second term has
the same meaning of the first one, but it is evaluated on the boundary of
the manifold. $\Psi $ is a trial wave functional of the gaussian form
restricted to the traceless-transverse sector (TT), which is gauge
invariant. No matter contribution has been considered. The one-loop
contribution shows that quantum fluctuations of the gravitational metric
shift the minimum of the effective energy from flat space ( the classical
minimum for the energy) to a multi-wormhole configuration. Indeed the total
energy contribution to one loop is
\begin{equation}
\Delta _{N_{w}}E\left( M\right) \sim -N_{w}^{2}\frac{V}{64\pi ^{2}}\frac{%
\Lambda ^{4}}{e},  \label{a1}
\end{equation}
where $V$ is the volume of the system, $\Lambda $ is the U.V. cut-off and $%
N_{w}$ is the wormholes number\cite{2,3,4}. This expression shows that a
non-trivial vacuum of the multi-wormhole type is favoured with respect to
flat space. It is important to remark that it is the $N$ - coherent
superposition of wormholes that it is privileged with respect to flat space
and not the single wormhole, because the single wormhole energy contribution
has an imaginary contribution in its spectrum: a clear sign of an
instability. Nevertheless the presence of an unstable mode is necessary to
have transition from one vacuum (the false one) to the other one (the true
vacuum)\cite{5}. Three consequences of this multiply connected spacetime are:

\begin{enumerate}
\item  the event horizon area of a black hole is quantized and by means of
the Bekenstein-Hawking relation\cite{6,7}, also the entropy of a black hole
is quantized. In particular for a Schwarzschild black hole, one gets
\begin{equation}
M=\frac{\sqrt{N}}{2l_{p}}\sqrt{\frac{\ln 2}{\pi }},
\end{equation}
namely the black hole mass is quantized.

\item  A cosmological constant is induced by vacuum fluctuations as shown by
Eq. $\left( \ref{a1}\right) $ whose value is
\begin{equation}
\Lambda _{c}=\frac{\Lambda ^{4}l_{p}^{2}}{N_{w}8e\pi }.  \label{a2}
\end{equation}
When the area-entropy relation is applied to the de Sitter geometry, we
obtain
\begin{equation}
\frac{3\pi }{l_{p}^{2}\ln 2N_{w}}=\Lambda _{c}.  \label{a3}
\end{equation}

\item  Combining Eq. $\left( \ref{a2}\right) $ and Eq. $\left( \ref{a3}%
\right) $, one gets
\begin{equation}
\Lambda _{c}=\frac{\Lambda ^{4}l_{p}^{2}}{N_{w}8e\pi }=\frac{3\pi }{%
l_{p}^{2}\ln 2N_{w}}.
\end{equation}
This means that we have found a constraint on the U.V. cut-off
\begin{equation}
\Lambda ^{4}=\frac{24e\pi ^{2}}{\ln 2l_{p}^{4}}.  \label{a4}
\end{equation}
This last consequence, although very encouraging, because at first glance
the model seems to be U.V. finite,  needs a very careful examination and
interpretation. In fact, since we have adopted a certain number of
approximations, Eq. $\left( \ref{a4}\right) $ could be only an artifact of
the calculation scheme. It is interesting to note that Eq. $\left( \ref{a1}%
\right) $ can be obtained at least for $N_{w}=1$ even for
Schwarzschild-Anti-de Sitter wormholes\cite{8}. In this case, the
Hamiltonian is
\begin{equation}
\Delta E\left( M,b\right) =E\left( M,b\right) -E\left( b\right) =\frac{%
\left\langle \Psi \left| H_{\Sigma }^{S-AdS}-H_{\Sigma }^{AdS}\right| \Psi
\right\rangle }{\left\langle \Psi |\Psi \right\rangle }+\frac{\left\langle
\Psi \left| H_{ql}\right| \Psi \right\rangle }{\left\langle \Psi |\Psi
\right\rangle },
\end{equation}
where $b=\sqrt{-3/\Lambda _{c}}$, the first term represents the difference
of the Hamiltonians evaluated on the Schwarzschild-Anti-de Sitter and
Anti-de Sitter spacetime respectively and the second term always represents
a subtraction procedure on the boundary of $\Sigma $. This means that a
selection rule has to emerge to compare the quantity
\begin{equation}
\Gamma _{{\rm N-S-AdS\ holes}}=\frac{P_{{\rm N-S-AdS\ holes}}}{P_{{\rm AdS}}}%
\simeq \frac{P_{{\rm foam}}}{P_{{\rm AdS}}}
\end{equation}
with
\begin{equation}
\Gamma _{{\rm N-holes}}=\frac{P_{{\rm N-holes}}}{P_{{\rm flat}}}\simeq \frac{%
P_{{\rm foam}}}{P_{{\rm flat}}}.  \label{a5}
\end{equation}
In both cases, we find a non-vanishing probability that a non-trivial vacuum
has to be considered. However in the case of Eq. $\left( \ref{a5}\right) $,
we have obtained that\cite{2}
\begin{equation}
\Gamma _{{\rm N-holes}}=P\sim \left| \exp \left( \frac{\Lambda _{c}}{8\pi
l_{p}^{2}}V_{c}\right) \left( \Delta t\right) \right| ^{2},
\end{equation}
which for the de Sitter case becomes
\begin{equation}
\exp \left( 3\pi /l_{p}^{2}\Lambda _{c}\right) .
\end{equation}
Thus we recover the Hawking result about the cosmological constant
approaching zero\cite{9}. Note that the vanishing of $\Lambda _{c}$ is
related to the growing of the wormholes number. Although these results are
quite encouraging, an important question comes into play: why a space-time
formed by wormholes has to be preferred with respect to the flat one, when
the last one is the one we observe. The answer that at this stage can only
be conjectured is that if we consider the following expectation value on the
foam state
\begin{equation}
\frac{\left\langle \Psi _{F}\left| \hat{g}_{ij}\right| \Psi
_{F}\right\rangle }{\left\langle \Psi _{F}|\Psi _{F}\right\rangle },
\end{equation}
when the number of wormholes is large enough, i.e. the scale is sufficiently
large, we should have to obtain
\begin{equation}
\frac{\left\langle \Psi _{F}\left| \hat{g}_{ij}\right| \Psi
_{F}\right\rangle }{\left\langle \Psi _{F}|\Psi _{F}\right\rangle }%
\rightarrow \eta _{ij},
\end{equation}
where $\eta _{ij}$ is the flat space metric. This is a test that this foamy
model has to pass if phenomenological aspects have to be considered.
\end{enumerate}

\bigskip
\textbf{Acknowledgments}

\smallskip
\noindent I would like to thank Prof. R. Ruffini who has given to
me the opportunity of participating to the Conference. I would
like also to thank Prof. A. Perdichizzi for a partial financial
support.


\bigskip

\begin{references}
\bibitem{1}  J.A. Wheeler, Ann. Phys. 2 (1957) 604; J.A. Wheeler,
Geometrodynamics. Academic Press, New York, 1962.

\bibitem{2}  R. Garattini, {\it A Spacetime Foam approach to the
cosmological constant and entropy}. To appear in Int.J.Mod.Phys.
D.;gr-qc/0003090.

\bibitem{3}  R. Garattini, Phys. Lett. B 446 (1999) 135, hep-th/9811187.

\bibitem{4}  R. Garattini, Phys. Lett. B 459 (1999) 461, hep-th/9906074.

\bibitem{5}  S. Coleman, Nucl. Phys. B 298 (1988) 178.

\bibitem{6}  J. Bekenstein, Phys. Rev. D7 (1973) 2333.

\bibitem{7}  S. Hawking, Phys. Lett. B 134 (1984) 403.

\bibitem{8}  R. Garattini, Class.Quantum Grav. 17 (2000) 3335, gr-qc/0006076.

\bibitem{9}  S. Hawking, Phys. Lett. {\bf B 134} (1984) 403.
\end{references}
\end{document}